\documentclass[twocolumn,a4paper,10pt]{article}

\usepackage[a4paper,top=2.2cm,bottom=2.2cm,left=1.8cm,right=1.8cm,
            columnsep=0.7cm]{geometry}
\usepackage{graphicx}
\usepackage{amsmath}
\usepackage{txfonts}
\usepackage{natbib}
\bibpunct{(}{)}{;}{a}{}{,}
\usepackage[colorlinks=true,allcolors=blue]{hyperref}

\newcommand{\degr}{\ensuremath{^{\circ}}}
\newcommand{\fdg}{\ensuremath{^{\circ}\mkern-7mu.\mkern1mu}}

\newcommand{\tablefoot}[1]{\par\vspace{3pt}{\footnotesize\textbf{Notes.} #1\par}}

\newcommand{\inst}[1]{\ensuremath{^{#1}}}
\newcommand{\email}[1]{\texttt{#1}}
\newcommand{\titlerunning}[1]{}
\newcommand{\authorrunning}[1]{}
\newenvironment{acknowledgements}%
  {\par\vspace{6pt}\noindent\small\textit{Acknowledgements.} }{\par}

\makeatletter
\newcommand{\institute}[1]{\gdef\@institute{#1}}
\gdef\@institute{}
\newcommand{\keywords}[1]{\gdef\@keywordlist{#1}}
\gdef\@keywordlist{}
\gdef\abstract#1#2#3#4#5{\gdef\@abstracttext{%
  \textbf{Context.} #1\ \textbf{Aims.} #2\ \textbf{Methods.} #3\
  \textbf{Results.} #4\ \textbf{Conclusions.} #5}}
\renewcommand{\maketitle}{%
  \twocolumn[\begin{@twocolumnfalse}
    \begin{center}
      {\LARGE\bfseries\@title\par}\vspace{10pt}
      {\large\@author\par}\vspace{6pt}
      {\footnotesize\@institute\par}
    \end{center}\vspace{8pt}
    {\small\@abstracttext\par\vspace{7pt}
     \noindent\textbf{Key words.} \@keywordlist\par}
    \vspace{8pt}\hrule\vspace{14pt}
  \end{@twocolumnfalse}]}
\makeatother

\usepackage[explicit]{titlesec}
\titleformat{\section}{\normalfont\large\bfseries}{\thesection.}{0.5em}{#1}
\titleformat{\subsection}{\normalfont\normalsize\bfseries}{\thesubsection.}{0.5em}{#1}
\titlespacing*{\section}{0pt}{12pt}{5pt}
\titlespacing*{\subsection}{0pt}{9pt}{3pt}

\begin{document}

\title{Stellar proper motions compared with the plane-of-sky magnetic field}

\author{S.~G. Ansari\inst{1}}

\institute{University of Vienna, Department of Astrophysics,
           T\"urkenschanzstrasse 17, A-1180 Vienna, Austria\\
           \email{ansaris22@univie.ac.at}}

\date{}

\titlerunning{The reference frame in proper-motion alignment statistics}
\authorrunning{S.~G. Ansari}

\abstract
{Several recent studies have compared the proper-motion directions of young stars
with the plane-of-sky magnetic field of their natal cloud, and reported preferred
relative orientations.}
{We ask how much of the reported orientation depends on the reference frame in which
the proper motions are expressed.}
{For young stellar objects in eight nearby clouds cross-matched to \textit{Gaia}~DR3
we compute the projected Rayleigh statistic against the \textit{Planck} 353\,GHz field
at each star's own position, in five frames: observed; the local standard of rest
(LSR); the cloud systemic motion removed; the same per kinematic subgroup, including
the perspective term; and with the fitted linear velocity gradient additionally
removed. Corona Australis is included as a control: it is the only one of these regions
for which the plane-of-sky motion of the gas has been measured directly.}
{The measured orientation changes by up to a factor of five and reverses sign between
frames; in the observed frame every cloud shows a strong preferred orientation. The
LSR frame is no improvement: in Orion\,A it triples the amplitude and flips its sign,
reaching $\langle\cos 2\Delta\theta\rangle = -0.84$. The tangential solar reflex
exceeds the internal velocity dispersion by factors of $2.6$--$18$, and depends on the
adopted solar motion at a level comparable to the correction itself. Removing the
systemic motion per subgroup brings all amplitudes below $0.28$; removing coherent
internal flows reduces them further, and in Corona Australis reverses their sign. We also
give a frame-independent bound, $|\langle\cos 2\Delta\theta\rangle| \le r_\theta r_\chi$,
which tests whether a measured amplitude contains any per-star information at all.}
{Any frame retaining a cloud's systemic motion returns the orientation of that motion,
not anything internal to the cloud. Measuring in the system's rest frame is standard in
cluster kinematics; we transfer this to field-alignment statistics and show the LSR
correction can be worse than none. How much further structure to remove is then a
modelling choice that changes the result and must be stated. The corrected frame is
that of the stars, not the gas, from which it differs, so the residual amplitudes are
upper limits; we make no claim about their physical origin.}


\maketitle

\section{Introduction}

\textit{Gaia} astrometry \citep{GaiaDR3} has made it practical to compare the
proper-motion directions of individual young stars with the plane-of-sky magnetic field
orientation of the cloud in which they formed. Several such comparisons have now been
published \citep{Sharma2022, Velguth2025}, reporting significant departures from random
relative orientation and reading them as a kinematic signature of the natal field.

Such a result does not survive a change of reference frame if the separations were computed
between the observed barycentric proper-motion vector and the local field. What such a
correlation would signify, if established, is contested,
and we do not enter that discussion here: once a star has decoupled from the gas the
interstellar field exerts no force on it, and nothing below should be read as suggesting
otherwise \citep[cf.][for feedback-driven cases]{Kounkel2020}. Our concern is prior to any
interpretation, namely whether the reported orientations measure anything internal to the
cloud at all.

They need not. The observed proper motion of a cloud member is dominated by the systemic
motion of the population --- solar reflex, differential Galactic rotation, and the cloud's
own space velocity --- whose direction is unrelated to any internal quantity. A statistic
built from proper-motion \emph{directions} in such a frame returns the orientation of the
systemic motion.

None of this is new as a matter of technique: in cluster kinematics internal motions are
measured in the rest frame of the system as a matter of course, with the perspective terms
subtracted along with the bulk motion \citep{vanLeeuwen2009, Kuhn2019,
VasilievBaumgardt2021}, and for the Orion complex the practice is standard
\citep{Grossschedl2021}. What has not been carried across is the application to
\emph{field-alignment} statistics, where the reference direction is external to the
population and the frame dependence is correspondingly easy to overlook. That transfer,
and the finding that the LSR frame --- the natural choice when no cluster rest frame is at
hand --- can leave a larger spurious signal than no correction at all, are what this
paper contributes.

\section{Data and method}

\subsection{Sample and field angles}

\begin{table*}[t]\centering
\caption{Adopted parameters for the nine regions. $d$ and $v_{r,\rm sys}$ are the values
used to build the L2/L3 rest frames and the perspective term. $r_{\rm fov}$ is the angular
radius spanned by the members about their centroid, and $\sigma_\chi$ the axial dispersion
of the \textit{Planck} 353\,GHz position angle over that footprint after smoothing to
$30'$, computed on doubled angles since $\chi$ is defined modulo $180\degr$. $N$ is the
number of \textit{Gaia}~DR3 sources surviving the cuts of this section.}
\label{tab:clouds}
\centering
\footnotesize
\setlength{\tabcolsep}{4pt}
\begin{tabular}{lrrrccl}
\hline\hline
Region & $N$ & $d$ & $v_{r,\rm sys}$ & $r_{\rm fov}$ & $\sigma_\chi$ & Membership \\
       &     & (pc) & (km\,s$^{-1}$) & (\degr) & (\degr) & catalogue \\
\hline
Chamaeleon\,I & 82   & 190 & $+14.5$ & 2.0  & 8  & \citealp{Esplin2017} \\
Ophiuchus     & 46   & 140 & $-7.0$  & 2.0  & 12 & \citealp{EsplinLuhman2020} \\
Orion\,A\,S   & 487  & 443 & $+27.0$ & 2.8  & 21 & \citealp{Grossschedl2018} \\
Orion\,A\,N   & 1000 & 400 & $+27.0$ & 2.0  & 9  & \citealp{Grossschedl2018} \\
Taurus        & 251  & 140 & $+16.5$ & 10.7 & 37 & \citealp{Luhman2023} \\
Lupus         & 47   & 160 & $+1.0$  & 6.3  & 22 & \citealp{Luhman2020} \\
IC\,348       & 78   & 321 & $+16.0$ & 0.3  & 5  & \citealp{Luhman2016} \\
NGC\,1333     & 31   & 293 & $+8.0$  & 0.2  & 2  & \citealp{Luhman2016} \\
CrA           & 324 &  150 & $+5.5$  & 5.8  & 34 & \citealp{Esplin2022} \\
\hline
\end{tabular}
\tablefoot{Total $N = 2346$. The systemic-motion vector subtracted at L2/L3 is formed from
$v_{r,\rm sys}$ and the sample mean proper motion at the centroid, taken per kinematic
subgroup where subgroups exist (Sect.~2.2); Taurus is subdivided into 13 such subgroups at
L3, so no single mean proper motion describes it. Orion\,A is split into N and S at
$\delta = -6\fdg3$. References for $d$: Chamaeleon\,I, \citealp{Dzib2018}, \citealp{Zucker2020};
Ophiuchus, \citealp{OrtizLeon2017}, \citealp{Zucker2019}; Orion\,A,
\citealp{Grossschedl2018}, \citealp{Kounkel2018}; Taurus, \citealp{Galli2019},
\citealp{Roccatagliata2020}; Lupus, \citealp{Galli2020}, \citealp{Dzib2018};
IC\,348 and NGC\,1333, \citealp{OrtizLeon2018a}.
Systemic velocities are heliocentric: Chamaeleon\,I, \citealp{Sacco2017};
Ophiuchus, \citealp{Rigliaco2016}; Orion\,A, \citealp{Kounkel2018},
\citealp{Grossschedl2021}; Taurus, member velocities from \citealp{Luhman2023};
Lupus, \citealp{Wichmann1999}. For IC\,348 and NGC\,1333 the systemic velocity is
determined here from the member sample of \citet{Luhman2016} rather than adopted, so
that it refers to the same stars from which the residuals are formed. For CrA both $d$ and $v_{r,\rm sys}$ are taken from \citet{Galli2020b}. CrA has no published kinematic substructure, so L2 and L3 coincide there exactly.}
\end{table*}

We use young stellar objects in eight nearby clouds (Chamaeleon\,I, Perseus split into
IC\,348 and NGC\,1333, Ophiuchus, Orion\,A South and North, Taurus, Lupus, Corona
Australis), assembled from
the membership catalogues listed in Table~\ref{tab:clouds} and cross-matched to
\textit{Gaia}~DR3 \citep{GaiaDR3}; Table~\ref{tab:clouds} also collects the adopted
distances, systemic radial velocities and footprint radii. We reject sources with
\texttt{RUWE}~$\ge 1.4$ \citep{Lindegren2021}, $\varpi/\sigma_\varpi < 5$, or
per-component proper-motion errors $>1$\,mas\,yr$^{-1}$, and apply a parallax window
around each cloud distance followed by a $3\sigma$ clip in proper-motion space. Orion\,A
is treated as two regions because it is not kinematically single --- the distance varies
along the cloud and the head differs in mean velocity from the tail
\citep{Grossschedl2018, Kounkel2018, Kounkel2022}, so one systemic motion would describe
neither well --- and we split at $\delta = -6\fdg 3$, being distance-based.

Corona Australis is included as a control rather than simply as a ninth region. It is
the only nearby star-forming region for which we are aware of a direct measurement of the
plane-of-sky motion of the gas \citep{Piecka2025}, and therefore the only one in which the
residuals of Sect.~3 can be referred to the frame of the gas rather than to that of the
stars --- the distinction on which Sect.~\ref{sec:gasframe} turns. We treat it throughout
as a control on that step, while reporting it in the tables alongside the rest.

Field orientations come from the \textit{Planck} 353\,GHz Stokes $Q,U$ maps
\citep{PlanckXXXV, Planck2020} smoothed to $30'$, with $\psi = \tfrac12\arctan2(-U,Q)$ and
$\chi = \psi + 90\degr$; the sign of $U$ was verified from the map header. As $Q,U$ are
defined in Galactic coordinates, $\chi$ is already a position angle east of Galactic
north. We sample it \emph{at each star's own position}: the field disperses by
$2$--$37\degr$ across these footprints, and a single cloud-mean angle misrepresents
individual sightlines by up to $89\degr$ (Taurus and Corona Australis, which both reach
the $90\degr$ ceiling that axial data impose). Because $\chi$ is an orientation
defined modulo $180\degr$, both quantities are computed on doubled angles; a linear
standard deviation is not a valid substitute, and would read $64\degr$ rather than
$37\degr$ for Taurus. The two Perseus clusters are the limiting case in the other
direction: their footprints, $0\fdg3$ and $0\fdg2$ in radius, are smaller than the
$30'$ smoothing beam, so per-star sampling there is equivalent to a single cloud-mean
angle and the choice is immaterial.

\subsection{The frame hierarchy}

We evaluate the same statistic in five frames. \emph{L0, observed}: barycentric proper
motions as catalogued. \emph{L1, LSR}: the solar reflex removed,
$\mu_{\rm LSR} = \mu_{\rm obs} + (\boldsymbol{v}_\odot\!\cdot\hat{\boldsymbol{e}})/(\kappa d)$
with $\kappa = 4.7405$\,km\,s$^{-1}$ per mas\,yr$^{-1}$\,kpc and
$\boldsymbol{v}_\odot = (11.10, 12.24, 7.25)$\,km\,s$^{-1}$ \citep{Schonrich2010}.
\emph{L2, cloud bulk removed}: one systemic motion per cloud, formed from the mean proper
motion and the systemic radial velocity at the centroid and re-projected to each star's
own position and distance before subtraction --- the perspective term, without which two
stars a degree apart moving identically retain different residual proper motions.
\emph{L3, subgroup bulk removed}: as L2 but per published kinematic subgroup (13 for
Taurus). A subgroup must retain at least five members for a mean to be subtracted from it,
which excludes three Taurus subgroups of four stars each; L3 and L4 there are evaluated on
the remaining $251$ of $263$ members, and each level is normalised by the number of stars
entering it.
\emph{L4, coherent flow removed}: L3 minus the fitted linear velocity gradient
$v = v_0 + \mathsf{G}\cdot\Delta x$.

The step from L0--L1 to L2--L4 is not a matter of degree: any frame retaining the systemic
motion measures a different quantity (Sect.~\ref{sec:why}). The choice among L2, L3 and L4
is a modelling decision (Sect.~\ref{sec:oversub}).

\subsection{Statistics}

For axial data we use the projected Rayleigh statistic \citep{Jow2018},
\begin{equation}
Z_x = \frac{\sum_i \cos 2(\theta_i - \theta_{{\rm ref},i})}{\sqrt{N/2}} ,
\end{equation}
where $\theta_i$ is the position angle of the proper motion of star $i$ and
$\theta_{{\rm ref},i}$ is the \textit{Planck} field position angle sampled at the sky
position of that same star (Sect.~2.1). No population-mean angle enters the statistic at
any point. We quote throughout the per-star amplitude
$\langle\cos 2\Delta\theta\rangle = Z_x/\sqrt{2N}$, which is bounded by unity and, unlike
$Z_x$, comparable between samples of different size.

Two further quantities are used in Sect.~\ref{sec:controls}. The \emph{preferred axis}
follows from $C = \sum_i\cos 2\theta_i/\sqrt{N/2}$ and $S = \sum_i\sin
2\theta_i/\sqrt{N/2}$, giving a reference-free strength $R = (C^2+S^2)^{1/2}$: the amount
of directional structure present without reference to any external direction. The
\emph{velocity dispersion tensor} is the $2\times2$ covariance of the Galactic velocity
components, with principal dispersions $\sigma_1 \ge \sigma_2$ and bootstrapped
uncertainties. For a zero-mean elliptical distribution the preferred axis and the
dispersion major axis coincide by construction, so comparing the observed $R$ with that
expected from $(\sigma_1,\sigma_2)$ tests whether a residual correlation is merely the
shape of the velocity ellipsoid.

\section{Results}

\begin{table}[t]\centering
\caption{Per-star amplitude $\langle\cos 2\Delta\theta\rangle$ between proper-motion
direction and the \textit{Planck} field, at each level of frame correction.}
\label{tab:main}
\centering
\footnotesize
\setlength{\tabcolsep}{3pt}
\begin{tabular}{@{}lrrrrrr@{}}
\hline\hline
Region & $N$ & L0 & L1 & L3 & L4 & $v_\odot^{\rm t}/\sigma$ \\
       &     & obs. & LSR & subgr. & flow & \\
\hline
Chamaeleon\,I & 82   & $+0.414$ & $-0.041$ & $+0.031$ & $+0.075$ & 18.2 \\
Ophiuchus     & 46   & $+0.670$ & $-0.286$ & $-0.167$ & $-0.175$ & 17.0 \\
Orion\,A\,S   & 487  & $+0.211$ & $-0.574$ & $-0.183$ & $-0.065$ & 2.6  \\
Orion\,A\,N   & 1000 & $+0.285$ & $\mathbf{-0.838}$ & $-0.159$ & $-0.172$ & 3.1 \\
Taurus        & 263  & $-0.331$ & $+0.224$ & $-0.050$ & $-0.057$ & 17.0 \\
Lupus         & 47   & $+0.653$ & $+0.744$ & $-0.223$ & $-0.110$ & 16.5 \\
IC\,348       & 78   & $+0.974$ & $+0.904$ & $+0.152$ & $+0.161$ & 15.5 \\
NGC\,1333     & 31   & $+0.972$ & $+0.424$ & $+0.273$ & $+0.333$ & 15.7 \\
CrA           & 324  & $+0.452$ & $+0.123$ & $-0.091$ & $+0.090$ & 9.1 \\
\hline
\end{tabular}
\tablefoot{$v_\odot^{\rm t}/\sigma$ is the tangential solar reflex divided by the internal
velocity dispersion: the factor by which the L1 correction exceeds the quantity being
measured. L2 is indistinguishable from L3 except in Taurus, the only region with
published subgroups, and is omitted for space. In Taurus L3 and L4 are evaluated on the
$251$ members lying in subgroups of at least five stars (Sect.~2.2); L0 and L1 use all
$263$, and each level is normalised by the number of stars entering it.}
\end{table}

\begin{figure}[t]\centering
\centering
\includegraphics[width=\columnwidth]{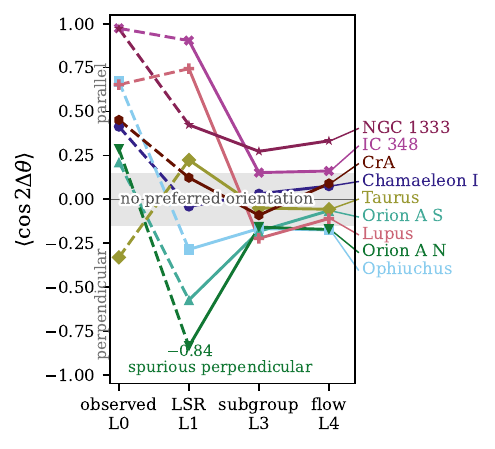}
\caption{Per-star amplitude between proper-motion direction and the \textit{Planck} field
at each level of frame correction. The grey band marks
$|\langle\cos 2\Delta\theta\rangle| < 0.15$.}
\label{fig:main}
\end{figure}

Table~\ref{tab:main} and Fig.~\ref{fig:main} give the per-star amplitude at each level of
correction.

\emph{In the observed frame the measured orientation is extreme.} Amplitudes reach $0.97$
in IC\,348 and NGC\,1333 --- values implying nearly every star oriented with the field ---
and every cloud shows a strong preferred orientation.

\emph{The LSR frame is not an improvement.} It reduces the amplitude in six regions,
increases it in three, and reverses its sign in five. In Orion\,A North it takes $+0.285$ to
$\mathbf{-0.838}$ ($Z_x = -37.5$): after ``correcting'' to the LSR, essentially every star
appears perpendicular to the field. The last column of Table~\ref{tab:main} shows why ---
the tangential solar reflex exceeds the internal dispersion by factors of $2.6$ to $18$,
so the transformation adds a coherent vector far larger than the internal motions. Where a
cloud's own space motion partly cancels the reflex in the observed frame, as in Orion,
removing the reflex leaves a \emph{larger} residual bulk motion than doing nothing.
Corona Australis is the one region where the LSR frame does neither harm nor much good:
the amplitude falls from $+0.452$ to $+0.123$ with the sign preserved, and its
$v_\odot^{\rm t}/\sigma = 9.1$ is the only value lying between $3.1$ and $15.5$ in
Table~\ref{tab:main}. The point is not that the correction is harmless there but that
nothing available beforehand distinguishes that case from Orion\,A North: the LSR step is
uncontrolled rather than systematically beneficial.

\emph{Removing the systemic motion brings everything into a narrow range}, with all
amplitudes falling to $|\langle\cos 2\Delta\theta\rangle| \le 0.28$.

\emph{Coherent internal flows survive that subtraction.} Removing the fitted gradient takes
the Lupus amplitude from $-0.223$ to $-0.110$, and the reference-free strength from
$R = 2.79$ to $0.53$: $81$ per cent of the Lupus value is an unremoved expansion of
$\delta = 0.17 \pm 0.06$\,km\,s$^{-1}$\,deg$^{-1}$. Orion\,A South loses $61$ per cent the
same way, and Corona Australis $66$ per cent ($R = 10.58 \rightarrow 3.62$), where the flow
is a near-pure expansion: $\delta = +0.26 \pm 0.01$\,km\,s$^{-1}$\,deg$^{-1}$ with a
rotational part $\omega = +0.01 \pm 0.01$ consistent with zero. There the subtraction does
not merely reduce the amplitude but reverses it, from $-0.091$ to $+0.090$, so that the
choice of level decides whether the population is reported as preferentially
perpendicular or parallel to the field.

\section{Discussion}

\subsection{Why a retained bulk motion dominates}
\label{sec:why}

Let a cloud have systemic tangential velocity $v_{\rm bulk}$ and internal dispersion
$\sigma_{\rm pec}$. Each star's observed direction scatters about the systemic direction by
roughly $\sigma_{\rm pec}/v_{\rm bulk}$, so the raw position angles are concentrated with
resultant $r$ and
\begin{equation}
Z_x \simeq \sqrt{2N}\; r \cos 2(\theta_{\rm bulk}-\theta_{\rm ref}) ,
\label{eq:artefact}
\end{equation}
the angle between the systemic motion and the reference direction scaled by $\sqrt{N}$: it
carries no information about the internal velocity field and becomes arbitrarily
significant in a large sample.

The same reasoning gives a bound that needs no frame at all. Write $r_\theta$ for the
resultant of the doubled proper-motion position angles and $r_\chi$ for that of the doubled
field angles; if the two are independent across stars then
\begin{equation}
|\langle\cos 2\Delta\theta\rangle| \;\le\; r_\theta\, r_\chi ,
\label{eq:ceiling}
\end{equation}
since $\langle\cos 2\Delta\theta\rangle = r_\theta r_\chi\cos 2(\bar\theta-\bar\chi)$ in
that case. Both factors are population-level quantities carrying no per-star information,
so an amplitude sitting at the bound carries none either: it is fixed by two mean
directions and their concentrations. The bound is worth reporting alongside the amplitude
whatever frame is adopted, because it separates a measurement that could reflect a
star-by-star relation from one that cannot.

The scope is set by the granularity of the statistic. Quantities built from velocity
\emph{differences} --- velocity structure functions, for instance --- are invariant under
the addition of a constant vector, and for those the frame is immaterial; the invariance
does not extend to statistics built from individual velocity \emph{directions}. This also
disposes of the intuition that the correction is unimportant for a compact region because
it is nearly constant across the footprint: compactness controls how much the correction
\emph{varies}, not its magnitude, which is set by the solar motion and the distance alone.
For a per-star directional statistic a uniform vector is the worst case rather than the
best, since it rotates every direction in the same sense.

The frame is therefore not one choice among equals. L0 and L1 are defined by the
observer's motion and bear no relation to the system under study; in them the statistic
returns a quantity that would be unchanged if the population had no internal velocity
structure at all. The stellar systemic frame is the only one in the hierarchy defined by
the members themselves (Sect.~\ref{sec:oversub}), and the dependence we document is
therefore removable, not irreducible.

The LSR frame carries a further difficulty. It makes the result depend on the adopted
solar motion, which is not settled: for Orion\,A North the correction is
$3.02$\,mas\,yr$^{-1}$ using \citet{Schonrich2010} but $1.61$\,mas\,yr$^{-1}$ using
\citet{DehnenBinney1998}, a difference comparable to the correction itself and driven by
the disputed $V$ component ($12.24$ versus $5.25$\,km\,s$^{-1}$). A frame defined by the
members themselves carries no such dependence.

That the LSR frame leaves a large shared vector behind is measurable independently of our
sample, and it agrees. For the Orion Nebula Cluster \citet{Kounkel2022} give a mean member
proper motion of $2.9$\,mas\,yr$^{-1}$ \emph{in the LSR frame} against an internal
dispersion of $0.9$--$1.1$\,mas\,yr$^{-1}$; restricted to the same footprint, our own L1
residuals for Orion\,A North give $3.0$ and $0.9$\,mas\,yr$^{-1}$. A frame that leaves
three dispersions of shared motion in place is not a rest frame.

\subsection{The frame is stellar, not the cloud's}
\label{sec:gasframe}

What L2 and L3 remove is the mean motion of the \emph{stars}: the frame they define is the
stellar systemic frame, not the rest frame of the gas. Direct measurement of the
plane-of-sky motion of interstellar material has recently become possible by tracking
cloud morphology across epochs, and in Corona Australis \citet{Piecka2025} find the gas
broadly comparable to the stars but systematically faster, with uncertainties they note
may be underestimated. The size of the offset is thus unsettled, but it need not be small
compared with the internal stellar dispersion --- the quantity out of which our residuals
are built.

The consequences are asymmetric. The argument against L0 and L1 (Sect.~\ref{sec:why}) is
untouched: it requires only that the retained systemic motion be large compared with the
internal dispersion, which holds whichever frame one regards as preferred. But the
residual \emph{amplitudes} at L2--L4 cannot be read as measurements of the stellar
velocity field relative to the gas, since an unknown bulk term of order the dispersion
itself may remain in them and would bias the residual directions coherently. We therefore
report them as upper limits and decline to interpret their magnitude or sign.

Residuals taken with respect to a directly measured gas velocity would settle this, and
Corona Australis is where that can now be done: it is the only region in
Table~\ref{tab:clouds} for which such a measurement exists \citep{Piecka2025}. Carrying
the residuals through in the gas frame is beyond the scope of this paper, and we note it
as the test the sample now admits rather than as a result claimed here. The criterion on which Corona Australis
was selected --- the availability of that measurement --- was fixed before any of its
statistics were computed.

\subsection{How much internal structure to remove}
\label{sec:oversub}

Removing structure can also remove the quantity of interest: if a cloud's subgroups are
parts of one system whose relative motions matter, subtracting a systemic motion from each
discards them, and the same applies at L4. L2, L3 and L4 therefore measure different
things, and we report all rather than nominating one. Taurus is the only region whose
published subgroups we use, and there
subtracting per subgroup \emph{increased} the preferred-axis strength
($R = 2.34 \rightarrow 5.11$), unmasking a within-group orientation rather than removing a
cloud-scale one. Orion\,A also has published substructure \citep{Kounkel2018, Kounkel2022}
which we do not use, and the coherent flows L4 removes there are physically established;
its L3 and L4 entries should be read with that in mind.

That increase is not an artefact of the subtraction itself. Removing a mean from each of 13
subgroups constrains the residuals --- each group is forced to sum to zero --- and could
raise $R$ without any physical structure being present. We tested this by permuting the
subgroup labels, preserving the 13 group sizes but randomising membership, and recomputing
L3. Over $500$ realisations the null distribution has a median of $2.02$, close to the L2
value, and does not exceed $3.17$; none reaches the observed $5.11$ ($p < 0.002$). The
structure unmasked at L3 is therefore in the data and not in the number of parameters
removed. We note that the converse case --- a rise consistent with the null --- would be
the more common one, and recommend the same test wherever per-subgroup subtraction is
applied to a small sample.

\subsection{The observed frame as a worked example}
\label{sec:v1}

The L0 column of Table~\ref{tab:main} is the clearest available illustration, because the
frame in which the statistic was evaluated is not in doubt. Every one of those nine values
would be read as a significant departure from random orientation, in the sense in which
such departures are usually reported, and every one is an orientation of the systemic
motion. The apparent variation between regions --- the feature that would invite an
environmental interpretation --- is the variation in the direction of each cloud's space
motion relative to its local field, which has no bearing on star formation.

\subsection{Other reported results}
\label{sec:others}

The same concern does not automatically apply elsewhere, and we have checked the nearest
case. \citet{Velguth2025} compare stellar proper motions with plane-of-sky field
orientations in Taurus and Perseus. The authors confirm (Y.~Li and B.~N.~Velguth, priv.\
comm.) that their alignment test was performed on residual velocities after subtracting
each group's systemic motion, which is the level we designate L3; the concern of
Sect.~\ref{sec:why} therefore does not apply to it. Since the step is not stated in their
text, we recommend only that the frame be reported explicitly in work of this kind, a
reader being otherwise unable to tell L0 from L3 --- a distinction worth up to a factor of
five in amplitude here.

One difference remains unresolved. \citet{Velguth2025} report that for their two Perseus
groups the subtraction made little difference, whereas our subgroup-corrected amplitudes
in IC\,348 and NGC\,1333 are $+0.152$ and $+0.273$ against raw values of $+0.974$ and
$+0.972$. The comparison is not like for like --- the studies differ in membership
catalogue, in the statistic applied to the angle distribution, and in the treatment of the
field, which we sample per star and which differs by $15\degr$ between IC\,348 and
NGC\,1333 despite their lying $3\degr$ apart --- and we record it as worth resolving
rather than adjudicating it here.

\subsection{Controls on any residual correlation}
\label{sec:controls}

A residual correlation surviving the frame correction is not thereby a measurement of the
field. Three of ours are not. In Orion\,A North what survives is an anisotropic velocity
ellipsoid, axis ratio $1.43 \pm 0.04$, whose major axis lies $2.8\degr$ from the cloud's
spatial long axis and which accounts for three quarters of the observed $R$: the
directions track the shape of the stellar distribution, which there happens to lie
near-perpendicular to the field. Corona Australis is the same effect in weaker form: the
dispersion major axis lies at $87\fdg2 \pm 12\fdg6$ against a spatial long axis of
$90\degr$, a separation of $2\fdg8$, and the observed $R$ exceeds that predicted from
$(\sigma_1,\sigma_2)$ by a factor $1.54$, so roughly two thirds of it is the shape of the
velocity ellipsoid. We record the alignment but not more than that: the axis ratio there,
$1.21 \pm 0.10$, is only $2.1\sigma$ from isotropic, so the anisotropy the alignment would
explain is itself barely established. In Lupus nothing survives. Separating a correlation with
the field from one with cloud geometry requires the controls of Sect.~2.3 together with a
demonstration that the field direction is distinguishable on the sky from the Galactic
plane and from the cloud's own elongation.

A second caveat is independent of the frame: in bound clusters the stellar velocities have
been reprocessed by the cluster potential, degrading any memory of the natal gas dynamics,
with the Orion Nebula Cluster the extreme case and IC\,348 not much better, while more
diffuse populations retain more \citep{Kounkel2018, Kounkel2022}. There the absence of a
residual correlation is weak evidence about initial conditions, and its presence weaker
still.

We therefore recommend: (i) evaluate directional statistics only in a frame defined by the
members themselves, including the perspective term; (ii) state which level of internal
structure has been removed, and report more than one; (iii) sample the field at each
star's position wherever the footprint exceeds the beam, and compute its dispersion on
doubled angles; (iv) test any surviving correlation against cloud elongation and the
Galactic plane before attributing it to the field; and (v) treat the amplitudes as upper
limits unless the gas velocity has been measured independently
(Sect.~\ref{sec:gasframe}); and (vi) quote $r_\theta$ and $r_\chi$ with the amplitude, so
that a reader can see from Eq.~(\ref{eq:ceiling}) whether the measurement contains any
per-star information at all.

\section{Conclusions}

\begin{enumerate}
\item A preferred orientation measured in the observed barycentric frame reflects the
      systemic motion of each cloud rather than anything internal to it, and is not
      evidence of a relationship between stellar motions and the natal field.
\item The measured orientation is set largely by the reference frame. Between the observed
      and subgroup-corrected frames the per-star amplitude changes by up to a factor of
      five and reverses sign. The invariance of pair-difference statistics under a constant
      offset does not extend to statistics built from individual velocity directions.
\item The LSR frame does not fix this and can make it worse, yielding
      $\langle\cos 2\Delta\theta\rangle = -0.84$ in Orion\,A North; the tangential solar
      reflex exceeds the internal dispersion by $2.6$--$18$ here.
\item Beyond removal of the systemic motion, how much internal structure to remove is a
      modelling choice that changes the result and must be stated. Coherent flows account
      for $81$ per cent of the Lupus value, $66$ per cent of that in Corona Australis and
      $61$ per cent of that in Orion\,A South; in Corona Australis their removal reverses
      the sign of the residual, so the modelling choice decides the reported sense of the
      alignment and not only its size.
\item A surviving correlation may still reflect the geometry of the stellar distribution:
      the Orion\,A North residual is a velocity ellipsoid aligned to $2.8\degr$ with the
      cloud's long axis, and the Corona Australis residual is a weaker instance of the
      same thing.
\item The corrected frame is that of the stars, not of the gas, and the two differ
      \citep{Piecka2025}. The case against L0 and L1 is unaffected, but the residual
      amplitudes are upper limits, not interpretable in magnitude or sign until the gas
      velocity is measured independently. Corona Australis, included here as a control on
      exactly this point, is the one region where that measurement now exists, and is
      where the test should be made.
\end{enumerate}

\begin{acknowledgements}
I thank S.~Meingast for comments on an earlier draft and for suggesting the Corona Australis comparison, M.~Kounkel for correspondence, and Y.~Li and B.~N.~Velguth for clarifying the procedure of their alignment analysis. This work has made use of data from the ESA mission \textit{Gaia}, processed by the \textit{Gaia} DPAC, and from \textit{Planck}, an ESA science mission. It made use of
Astropy, Matplotlib, healpy, and the SIMBAD and VizieR services operated at CDS, Strasbourg. The authors used Claude (Anthropic) to assist with data-analysis code, exploratory analysis, and drafting of the manuscript. All analysis was verified by the authors, and the authors take full responsibility for the content.
\end{acknowledgements}


\end{document}